\documentstyle[epsfig,epsf,psfig,aps,preprint]{revtex}
\newcommand{\be}{\begin{equation}}
\newcommand{\ee}{\end{equation}}
\newcommand{\bel}[1]{\be\label{#1}}
\newcommand{\re}[1]{Eq.~(\ref{#1})}
\newcommand{\mbs}[1]{\mbox{$\scriptstyle{#1}$}}
\newcommand{\ds}{\displaystyle}
\newcommand{\psib}{\overline{\psi}}
\newcommand{\goo}{\,\raisebox{-.5ex}{$\stackrel{>}{\scriptstyle\sim}$}\,}
\newcommand{\loo}{\,\raisebox{-.5ex}{$\stackrel{<}{\scriptstyle\sim}$}\,}
\newcommand{\dd}{\partial\hspace{-7pt}/}

\begin{document}
\renewcommand{\thefootnote}{\arabic{footnote}}

\title{Strange quark stars within the Nambu--Jona-Lasinio model}

\author{M. Hanauske$^1$, L.M. Satarov$^{1,2}$, I.N. Mishustin$^{1,2,3}$,
H. St\"ocker$^1$, and W. Greiner$^1$}

\address{$^1$Institut f\"ur Theoretische Physik,
J.\ W.\ Goethe--Universit\"at, D-60054 Frankfurt, Germany\\
$^2$The Kurchatov Institute, 123182 Moscow, Russia\\
$^3$The Niels Bohr Institute, DK-2100 Copenhagen O, Denmark
}

\maketitle

\begin{abstract}
We investigate the properties of charge--neutral $\beta$--equilibrium
cold quark matter within the Nambu--Jona-Lasinio model. The
calculations are carried out for different ratios of coupling constants
characterizing the vector and scalar 4--fermion interaction,
$\xi\equiv G_V/G_S$. It is shown that for $\xi < 0.4$ matter is self--bound
and for $\xi < 0.65$ it has a first order phase transition of the
liquid--gas type. The Gibbs conditions in the mixed phase are applied
for the case of two chemical potentials associated with the baryon
number and electric charge. The characteristics of the quark stars are
calculated for $\xi = 0, 0.5$ and $1$. It is found that the phase
transition leads to a strong density variation at the surface of these
stars. For $\xi = 1$ the properties of quark stars show behaviors
typical for neutron stars. At $\xi\goo 0.4$ the stars near to the
maximum mass have a large admixture of strange quarks in their
interiors.

PACS number: 14.65.-q, 26.60.+c, 97.10.-q 
\end{abstract}

\draft

\section{Introduction}

The direct application of QCD at moderate temperatures and nonzero
baryon densities is not possible at present. Therefore, more simple
effective models respecting some basic symmetry properties of QCD are
commonly used. The Nambu--Jona-Lasinio (NJL) model \cite{Nam61,Lar61}
which is dealing with constituent quarks and respects chiral symmetry
is one of the most popular model of this kind. In recent years this
model has been widely used for describing hadron properties (see
reviews \cite{Vog91,Kle92}), phase transitions in dense matter
\cite{Asa89,Kli90a,Kle99} and multiparticle bound states
\cite{Bub98,Mis99,Mis00}. This model has been also applied for studying
the Equation of State (EoS) of $\beta$--equilibrated matter appropriate
for stars \cite{Scha99,Stei00}. Unfortunately, the repulsive vector
interaction, which is very important in dense baryon--rich environment
\cite{Kli90a,Mis99}, was not taken into account there. The main goal of
our paper is to study the influence of this interaction on the EoS and
the structure of quark stars. We assume that these stars are composed
of pure quark matter, similar to previous studies using a simplified
EoS of the MIT bag model \cite{Weber,Pes00}. We do not consider hybrid
stars where the quark matter is matched to the hadron matter at low
densities. The reason of this is two--fold. First, all matching
procedures are quite ambiguous and model dependent. Second, our
calculations show that the interiors of such stars have so high
densities that the quark degrees of freedom seem to be more
appropriate. At least this should be a reasonable first approximation
to hybrid stars. One more advantage of using quark degrees of freedom
is related to the problem of strangeness content of the star matter. In
recent years many calculations have been done in hadronic models where
strangeness appears through either hyperon formation \cite{SM96} or
kaon condensation \cite{Brown94,Pons00,Ramos00}. Within the NJL model the
strangeness degree of freedom is handled in a very simple way through
the population of matter by strange quarks at sufficiently high baryon
densities.

\section{$\beta$--equilibrium quark matter}
\subsection{Description of the model}
Below we use the SU(3)--flavour version of the NJL model
proposed in Ref.~\cite{Reh96}, but including not only the scalar
but also the vector 4--fermion interaction. The color
singlet part of the Lagrangian in the mean field
approximation can be written as \cite{Mis00}
($\hbar=c=1$)
\begin{eqnarray}
{\cal L}&=&\sum_f\psib_f\,(i\dd-m_f-\gamma_0 G_V\rho_f)\,\psi_f
-\frac{\ds G_S}{\ds 2}\sum_f\rho_{\mbs{Sf}}^{\,2}\nonumber\\
&+&\frac{\ds G_V}{\ds 2}\sum_f\rho_f^{\,2}
+4K\prod_f\rho_{\mbs{Sf}}+{\cal L}_e\,.\label{lagrm}
\end{eqnarray}
Here~$\psi_f$~is~the~field operator of quarks with
flavour \mbox{$f=u,d,s$} and
\begin{eqnarray}
\rho_{\mbs{Sf}}&=&<\psib_f\psi_f>\,,\label{dens}\\
\rho_{\mbs{f}}&=&<\psib_f\gamma_0\psi_f>\label{denv}
\end{eqnarray}
are their scalar and vector densities.
Angular brackets in Eqs.~(\ref{dens})--(\ref{denv})
denote quantum--statistical averaging.
$G_S$, $G_V$ and $K$ in \re{lagrm} are respectively the coupling
constants of scalar, vector and flavour--mixing interactions.

The last term in \re{lagrm} is the leptonic part of the Lagrangian.
Below we take into account only electrons and treat them as ideal gas
of massless particles. As calculation show, the maximum value of the
electron chemical potential $\mu_e$ does not exceed significantly the
muon mass so that the muon admixture should be small. In this
approximation
\bel{llagr}
{\cal L}_e=\psib_e\,i\dd\,\psi_e\,,
\ee
where $\psi_e$ is the electron field.

The constituent quark masses, $m_f$, are determined from the coupled set of
gap equations
\bel{gape}
m_f=m_{0f}-G_S\,\rho_{\mbs{Sf}}+
2K\,\prod_{f'\neq f}\rho_{\mbs{Sf'}}\,,
\ee
where $m_{0f}$ is the bare (current) mass of quarks with
flavour $f$\,. The single--particle energies of quarks are equal
to $E_f(p)=\sqrt{m_f^{\,2}+p^2}$\,.

Within this model chiral condensates $<\psib_f\psi_f>$ are given by the
scalar densities of quarks occupying both the negative and positive
energy states. The divergent contributions of negative energy states
of the Dirac sea are regularized by introducing a 3--momentum cut--off
$\Theta(\Lambda-p)$ in momentum space integrals. In the case of
spatially homogeneous matter at zero temperature one arrives at the
following expression \cite{Mis99}
\bel{scald}
\rho_{\mbs{Sf}}=
-\frac{\ds\nu m_f}{\ds 2\pi^2}\int\limits_{p_{\mbs{Ff}}}^\Lambda
\frac{{\rm d}p\,p^2}{E_f(p)}=
\frac{\ds\nu m_f}{\ds 4\pi^2}\,
\left[\,p_{\mbs{Ff}}^{\,2}\,\Phi\left(\frac{\ds m_f}
{\ds\,p_{\mbs{Ff}}}\right)
-\Lambda^2\Phi\left(\frac{\ds m_f}{\ds\Lambda}\right)\right]\,.
\ee
Here $\nu=6$ is the spin--color degeneracy
factor,
$p_{\mbs{Ff}}=(\frac{\ds 6\pi^2}{\ds\nu}\rho_f)^{1/3}$
is the Fermi momentum of quarks with flavour $f$ and
\bel{phif}
\Phi(x)=\sqrt{1+x^2}-\frac{\ds x^2}{\ds
2}\ln{\,\frac{\sqrt{1+x^2}+1}
{\sqrt{1+x^2}-1}}\,.
\ee
The model parameters $m_{0f}, G_S, K, \Lambda$ can be fixed by
reproducing the observed masses of $\pi, K$\,, and $\eta'$ mesons
as well as the pion decay constant $f_\pi$.
As shown in Ref.~\cite{Reh96},
a reasonable fit is achieved with the following values:
\begin{eqnarray}
m_{0u}=m_{0d}&=&5.5~{\rm MeV},~~~m_{0s}=140.7~{\rm
MeV},\\
G_S=20.23~{\rm GeV}^{-2}, &&K=155.9~{\rm GeV}^{-5},
~~~\Lambda=0.6023~{\rm GeV}.
\label{papa}
\end{eqnarray}

Several attempts have been made to extract the vector coupling constant
by fitting the nucleon axial charge $g_A$~\cite{Vog91} and masses of
vector mesons~\cite{Pol97}. It was established that the ratio of the
vector and scalar coupling constants $\xi\equiv G_V/G_S$ should be of
the order of unity.  But still there is no agreement between the values
used by different authors. As demonstrated in Refs.~\cite{Mis99,Mis00},
the EoS of baryon--rich matter, in particular, the existence of bound
states and phase transitions is highly sensitive to $\xi$\,. Due to
uncertainty in the parameter $G_V$\,, below we present the results for
various values of $\xi$ within the interval $0\le\xi\le 1$\,.

\subsection{Equation of state}
We assume that the quark matter is in chemical
equilibrium with respect to the strong and weak interactions.
If neutrinos accompanying weak processes escape freely
($\mu_\nu=0$),
one obtains the following conditions
\bel{cheme}
\mu_d=\mu_s=\mu_u+\mu_e\,,
\ee
relating the chemical potentials $\mu_i$ of various particle species
$i=u,d,s,e$\,.
These equations are automatically satisfied
if $\mu_i$ are represented as linear combinations of the
baryon~($\mu_B$), strange~($\mu_S$) and charge ($\mu_Q$) chemical potentials:
\bel{chemi}
\mu_i=B_i\mu_B+S_i\mu_S+Q_i\mu_Q\,,
\ee
where $B_i$, $S_i$ and $Q_i$ are the baryon, strangeness
and charge quantum numbers of the particle species $i$\,. From
Eqs. (\ref{cheme}) it is evident that $\mu_S = 0$ and $\mu_Q=-\mu_e$.
In the considered case of zero temperature
the chemical potentials can be written in the explicit
form:
\begin{eqnarray}
\mu_f&=&\sqrt{m_f^{\,2}+p_{\mbs{Ff}}^2}+G_V\rho_f\,,\label{muf}\\
\mu_e&\simeq &p_{\mbs{Fe}}=(3\pi^2\rho_e)^{1/3}\,,\label{mue}
\end{eqnarray}
where $\rho_e$ is the number density of electrons.

The chemical potentials, constituent quark masses
and particle densities $\rho_i$ are found by simultaneously solving
Eqs.~(\ref{gape})--(\ref{scald}),~(\ref{cheme})--(\ref{mue})
under conditions of fixed baryon and charge densities,
\begin{eqnarray}
\rho_B&=&\sum_i
B_i\rho_i=\frac{1}{3}\,(\rho_u+\rho_d+\rho_s)\,,\\
\rho_Q&=&\sum_i Q_i\rho_i=
\frac{1}{3}\,(2\rho_u-\rho_d-\rho_s)-\rho_e\,.
\end{eqnarray}
Unless stated otherwise, the condition of local charge neutrality
$\rho_Q=0$ is assumed.

To obtain the EoS of quark matter,
one can calculate the energy density $\epsilon=T^{00}$
directly from the Lagrangian~(\ref{lagrm}). At zero
temperature one arrives at the expression (see
Ref.~\cite{Mis00} for details)
\begin{eqnarray}
\epsilon&=&\sum_f
\left[\,\frac{\ds\nu}{\ds 2\pi^2}\int\limits_{p_{\mbs{Ff}}}^\Lambda
{\rm d}p\,p^2 E_f(p)+\frac{\ds G_S}{\ds 2}\rho_{\mbs{Sf}}^{\,2}
+\frac{\ds G_V}{\ds 2}\rho_f^{\,2}\right]\nonumber\\
&-&4K\prod_f\rho_{\mbs{Sf}}+\frac{\ds p_{\mbs{Fe}}^4}{4\pi^2}
-\epsilon_{\rm vac}\,.\label{ende}
\end{eqnarray}
The constant $\epsilon_{\rm vac}$ is introduced in
order to set the energy density of the
physical vacuum ($p_{\mbs{Fi}}=0$) equal to zero. This constant
can be expressed through the vacuum
values of constituent quark masses. The latter are obtained by
solving the gap equations (\ref{gape})
for the case $p_{Ff}=0$\,.
At zero temperature the pressure can be obtained from the
energy density by using the thermodynamic identity
\bel{pres}
P=\sum\limits_i\mu_i\rho_i-\epsilon\,.
\ee

To characterize the flavour composition of quark matter, we
introduce the strangeness fraction parameter
\bel{rs}
r_s=\frac{\ds |S|}{3B}=\frac{\ds\rho_s}{\ds \rho_u + \rho_d + \rho_s}\,
\ee
where $B$ and $S$ are the baryon charge and strangeness of matter.

\subsection{Possibility of a phase transition}\label{sec2c}
\begin{figure}[ht]
\centerline{\mbox{
\epsfxsize=12cm\epsffile{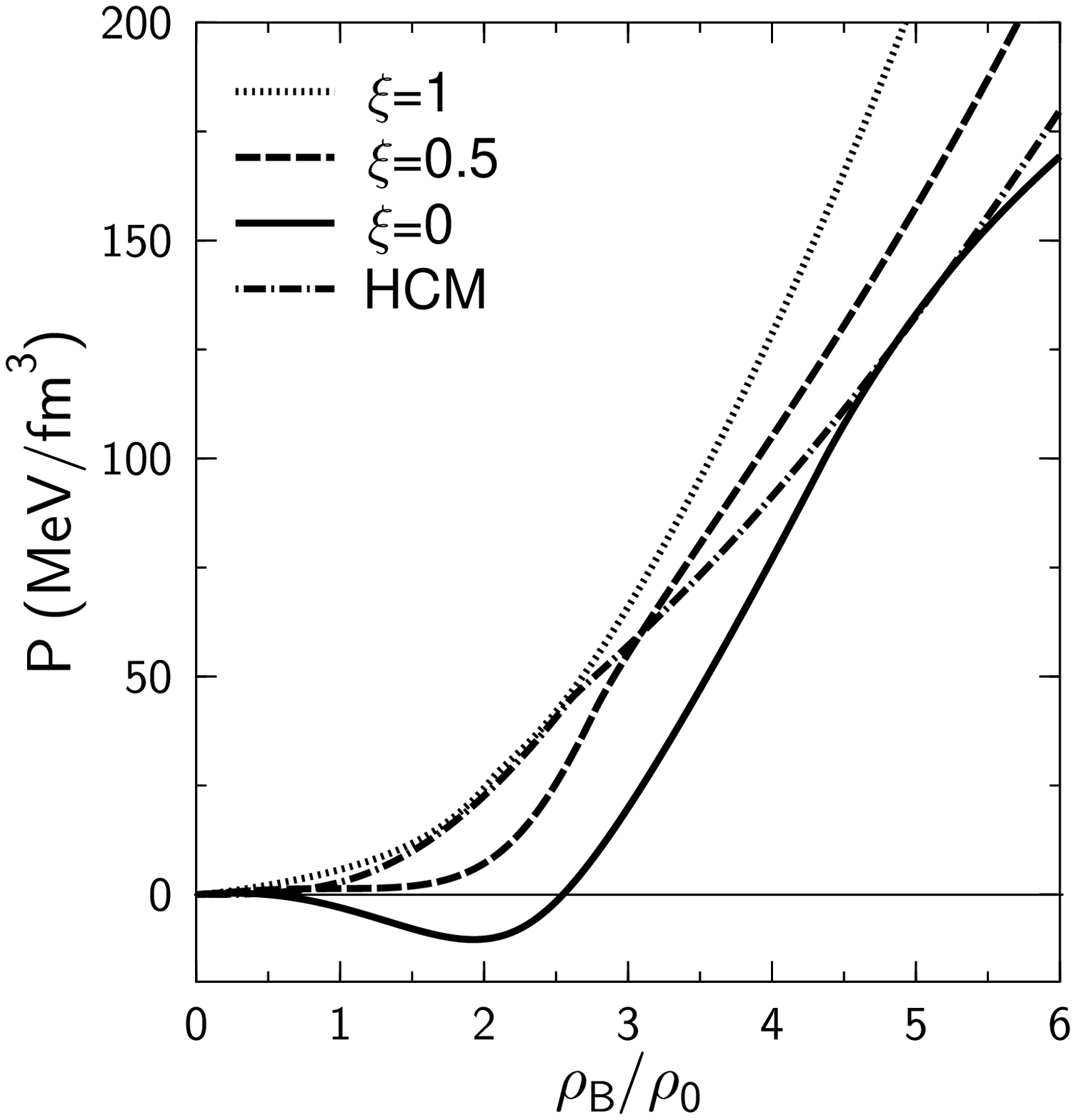}}}
\caption{Pressure as a function of the baryon density (in units of
$\rho_0=0.15$~fm$^{-3}$) calculated within the NJL model for
different ratios of vector and scalar coupling constants $\xi$\,. In
the case of $\xi = 0$ the matter is self--bound that is signaled by the
presence of negative pressures. Dashed--dotted line shows
the predictions of the HCM (with hyperons).} \label{fig_1}
\end{figure}

\begin{figure}[ht]
\centerline{\mbox{
\epsfxsize=12cm\epsffile{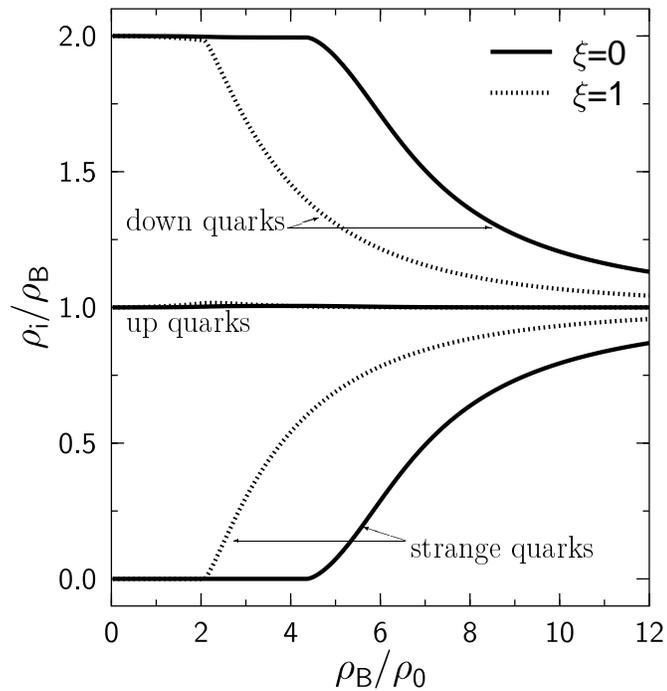}}}
\caption{Quark flavour abundances $\rho_f/\rho_B$ (f=u, d, s) versus
the baryon density $\rho_B$ for two different ratios of the vector
and scalar coupling constants $\xi = 0$ and $1$.} \label{fig_2}
\end{figure}

Some results concerning properties of the $\beta$--equilibrated quark
matter at zero temperature are presented in Figs. \ref{fig_1} and
\ref{fig_2}. Fig.~\ref{fig_1} shows pressure as a function of baryon
density for three values of the parameter $\xi$, namely for $\xi=0,\,0.5$ and 1.
Below we compare our results with the predictions of the Hadron Chiral
Model (HCM) \cite{Hana00,Papa99} for the $\beta$--equilibrated hadronic
matter. Pressure calculated in that model is also shown in
Fig.~\ref{fig_1}. Fig.~\ref{fig_2} represents equilibrium
concentrations of different quark flavours in $\beta$--equilibrium
matter. Comparison of the NJL model predictions for different $\xi$
reveals a strong sensitivity to this parameter\,. At $\xi=0$ one can
see two zero--pressure points at nonzero baryon densities.  They
correspond to a local maximum and a local minimum in the energy per
baryon $\epsilon/\rho_B$\,. In the considered case the local minimum
corresponds to a bound state of quark matter. This means that finite
droplets of such matter can exist in mechanical equilibrium with vacuum
even without gravitational force.

At $\xi\sim 0.5$ the pressure curve $P=P(\rho_B)$ does not cross zero,
but still contains parts  unstable with respect to the baryon and
charge density  fluctuations. This implies the possibility of a first
order phase transition of the liquid--gas type. In the case of
isospin--symmetric matter similar phase transitions have been studied
in Ref.~\cite{Mis00,Mis00a}\,. As demonstrated in Fig.~3 parameters of
the phase transition are rather sensitive to $\xi$\,. Dashed lines in
this figure show the states unstable with respect to the decomposition
of matter into two ($k=1,2$) coexisting phases with different baryon
($\rho_{B}^{(k)}$) and charge ($\rho_{\rm Q}^{(k)}$) densities. The
spatially averaged baryon density is defined as
\bel{abrd}
\rho_B=\lambda\,\rho_{B}^{(1)}+(1-\lambda)\,\rho_{B}^{(2)}\,,
\ee
where $\lambda$ is the volume fraction occupied by the denser phase
($0\le\lambda\le 1$)\,.

\begin{figure}[ht]
\centerline{\mbox{
\epsfxsize=12cm\epsffile{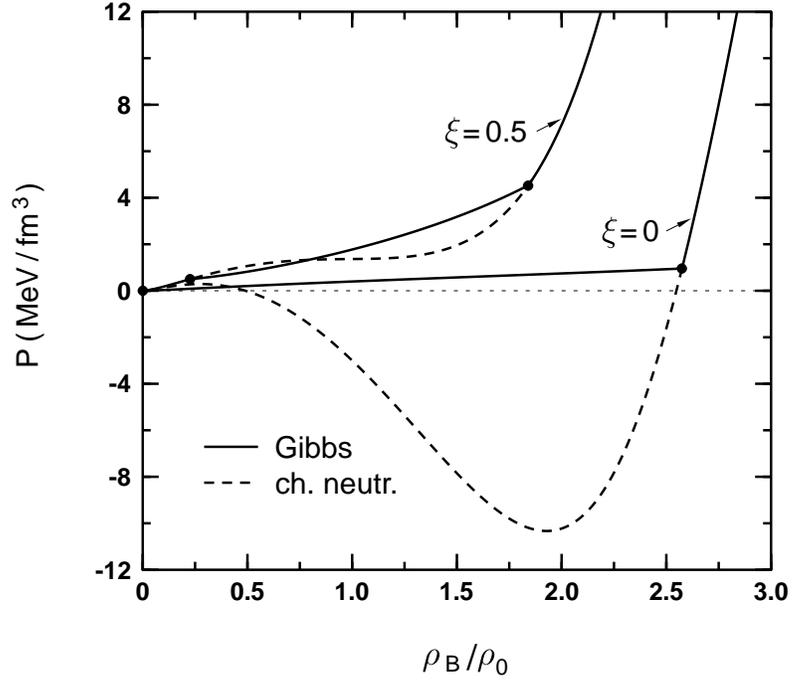}}}
\caption{Pressure as a function of the baryon density for $\xi = 0$ and
$0.5$ when a phase transition of the liquid--gas type is predicted by
the NJL model. The solid lines between the dots correspond to the mixed
phase obtained by applying the Gibbs conditions.  Dashed lines show the
results of calculations without the mixed phase.} \label{fig_3}
\end{figure}

The Gibbs conditions
of two--phase equilibrium, relating pressures
$P^{(k)}=P\left(\rho_{B}^{(k)},\rho_Q^{(k)}\right)$,
and chemical potentials of the two phases can be written as
\begin{eqnarray}
P^{(1)}&=&P^{(2)}\,,\\
\mu_B^{(1)}&=&\mu_B^{(2)}\,,\\
\mu_Q^{(1)}&=&\mu_Q^{(2)}\,.
\end{eqnarray}
In accordance with the general discussion of Ref.~\cite{Gle92},
these conditions can be fulfilled simultaneously only at nonzero
$\rho_Q^{(k)}$\,. As a consequence, only the
``global'' charge neutrality condition
\bel{glcn}
\lambda\,\rho_Q^{(1)}+(1-\lambda)\,\rho_Q^{(2)}=0
\ee
should be satisfied in the mixed phase region. By using
Eqs.~(\ref{abrd})--(\ref{glcn}) one can calculate
the equilibrium pressure in this region as a function of
baryon density $\rho_B$\,.
As shown in Fig.~\ref{fig_3} this pressure increases nearly linearly with
$\rho_B$\,. Deviations from the Maxwell construction
($P={\rm const}$) are more visible at larger $\xi$\,.
However, the mixed phase region becomes more
narrow in this case and disappears completely at $\xi\simeq
0.65$\,. At $\xi=1$ no traces of this phase transition are
present any more and the resulting EoS is rather stiff.

\section{Properties of quark stars}
In this section the above derived EoS is used to construct the
models of compact stars composed of pure quark matter.  As will be
shown below, the properties of such stars differ significantly
depending on whether the matter is self--bound or not. In the case of
self--bound electrically--neutral matter one would expect the existence
of macroscopic objects of any size above a certain critical mass
determined by the surface effects.  But certainly the gravitational
interaction becomes more and more important when the mass of these
objects grows.

The gravitational field is described in a standard way by using a
spacetime--dependent metric $g_{\mu\nu}(x^\alpha)$ obeying the
Einstein's equations \cite{MTW}.  We consider only spherically
symmetric configurations and neglect all dynamical effects, like
oscillation and rotation. Moreover, we restrict our considerations to
stars without magnetic field and at zero temperature. It is also
assumed that matter can be treated as an ideal fluid. Under these
assumptions the Einstein's equations are reduced to the
Tolman--Oppenheimer--Volkoff (TOV) equations \cite{Tolm39}. For a given
EoS and a fixed central baryon density $\rho_c$ the inside solution of
the metric $g_{\mu\nu}(r)$, the energy density and pressure profiles
can be determined by solving the TOV equations until
the radius $R$ where the pressure vanishes.
The outside solution is given by the Schwarzschild metric and
depends only on the total gravitational mass $M$ of the star.

The low--density ($\epsilon < \epsilon_{\rm drip}= 4.3 \cdot 10^{11}$~g/cm$^3$)
outer layer of a star ('crust') contains mainly
nuclei and electrons. To describe this nuclear crust we use the
EoS suggested in Ref.~\cite{bay71a}.

\subsection{Mass--density curves}
\begin{figure}[ht]
\centerline{\mbox{
\epsfxsize=12cm\epsffile{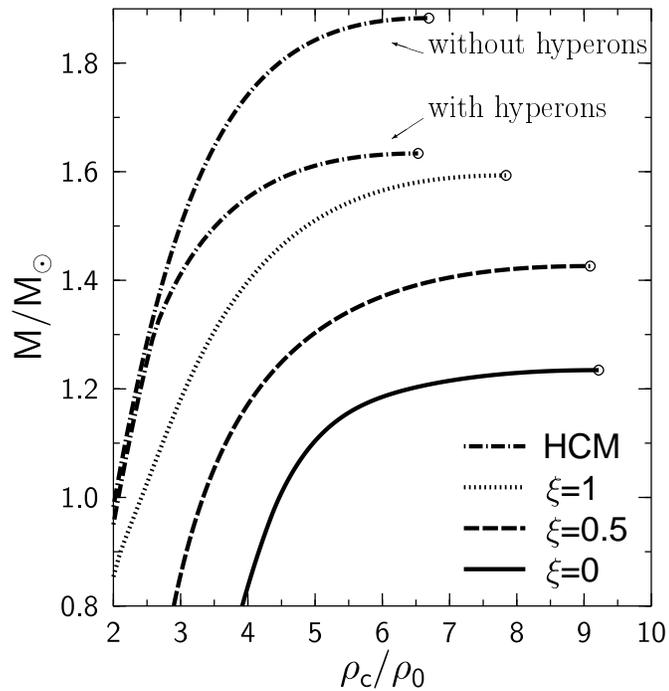}}}
\caption{Gravitational mass of a star M in units of the solar mass
$M_{\odot}$ versus the central baryon density $\rho_c$.  The three
lower curves show the predictions for quark stars within the NJL
model for different values of parameter $\xi$. The dashed--dotted
curves correspond to predictions of the HCM with and without hyperons.}
\label{fig_4}
\end{figure}

Fig. \ref{fig_4} represents the gravitational mass of stars as a
function of central baryon density~$\rho_c$. One can see that the
maximum masses predicted by the NJL model depend strongly on the
relative strength of the scalar and vector interactions $\xi$.
At $\xi= 0$ the maximum mass has a quite low value
$M_{\rm max} =  1.23 \,M_\odot$
with a high central baryon density
$\rho_c^{{\rm max}} = 9.2\, \rho_0$,
which is typical for quark matter stars\cite{Weber,glen}.
For higher values of $\xi$ the EoS becomes stiffer (see Fig.
\ref{fig_1}) which results in increasing maximum mass. On the other
hand, the central density $\rho_c^{{\rm max}}$ of the maximum--mass
stars decreases with $\xi$.  For $\xi = 1$ the maximum--mass star has
$M_{\rm max} = 1.60 \, M_\odot$ and $\rho_c^{{\rm max}} = 7.8 \, \rho_0$.

In Fig. \ref{fig_4} our results are compared with the predictions for
neutron stars obtained within the HCM \footnote{The slight differences
between the results presented here and in \cite{Hana00} are due to
improved numerics.}. The upper curve corresponds to the HCM calculation
where hyperons are neglected whereas the lower one shows the results
with the inclusion of hyperons. One can see that at higher $\xi$ the
NJL results for quark stars become close to the predictions of the HCM
(with inclusion of hyperons) for neutron stars.

\subsection{Mass--radius relations}
\begin{figure}[ht]
\centerline{\mbox{
\epsfxsize=12cm\epsffile{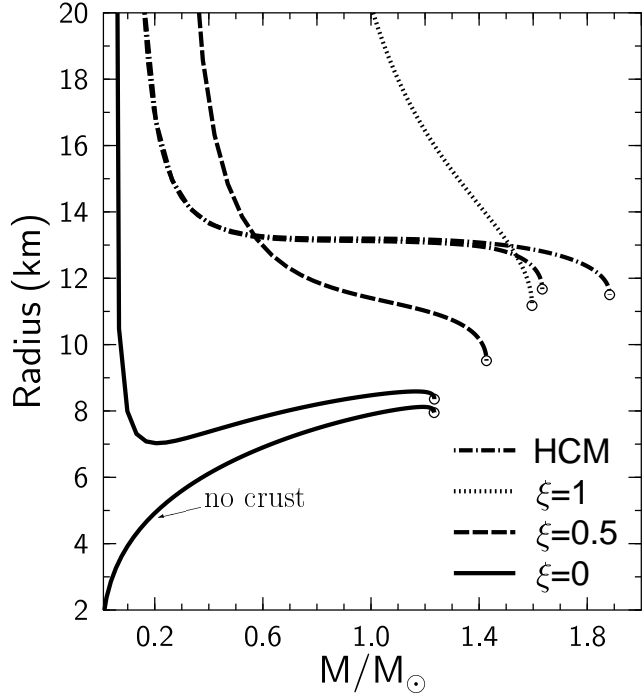}}}
\caption{Mass--radius curves predicted by the NJL model for three
values of the parameter $\xi$\,. For comparison, the predictions of
the HCM with and without hyperons are also shown. The results for
self--bound stars (solid lines) are presented for two calculations with
and without crust.} \label{fig_5}
\end{figure}

The calculated mass--radius relations are presented in Fig.
\ref{fig_5}. Again the predictions of the NJL model are compared with
the results of the HCM. One can see that at $\xi = 1$ the value of the
minimal radius $R_{\rm min} = R\,(M_{\rm max}) \simeq 11.2$ km predicted
by the NJL model is close to the HCM prediction. Here, again we
conclude that at large $\xi$ the properties of quark and neutron stars
are similar.

At decreasing $\xi$ the minimal radius becomes closer to the
predictions for quark stars made in Refs. \cite{Weber,glen}.  For
self--bound matter ($\xi < 0.4$) the mass--radius relation changes in a
qualitative way. In this case the corresponding curves start from the
origin and $M \propto R^3$ at small $R$ (without crust). It is not
clear to us whether it is necessary to include the crust for these
self--bound stars or not. But if we do this following the standard
prescriptions \cite{Weber} we see that the low--mass stars acquire an
extended mantel of crust, which may reach hundreds of kilometers. On
the other hand for quark stars with high masses (near to the maximum
mass) the inclusion of the crust leads to a relatively small increase
of the radii, of about several hundred meters.

\subsection{Density profiles}
\begin{figure}[ht]
\centerline{\mbox{
\epsfxsize=12cm\epsffile{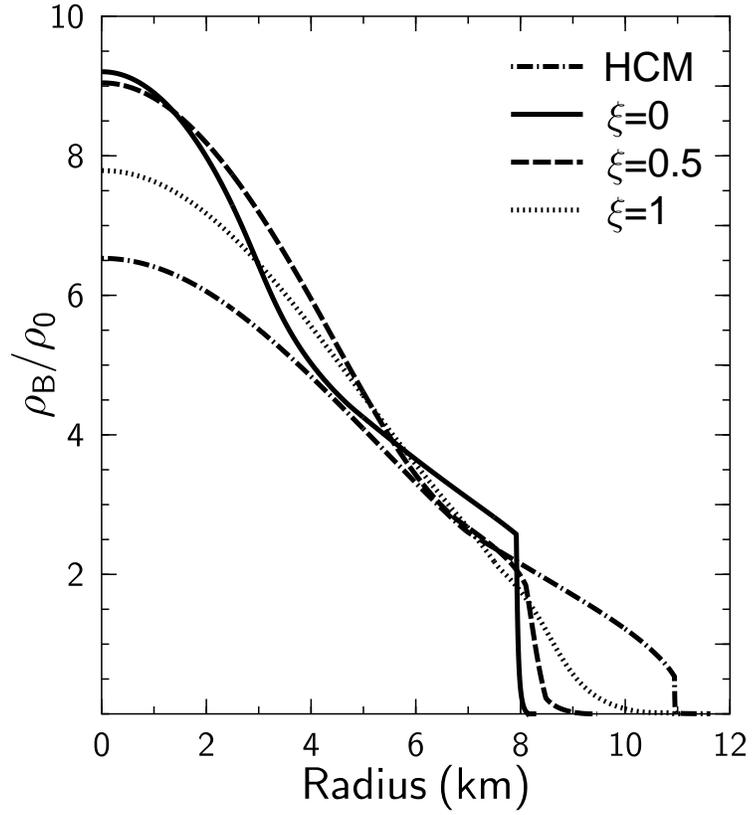}}}
\caption{Baryon density profiles for the maximum--mass quark stars
predicted by the NJL model for three different values of the parameter
$\xi$\,. The dashed--dotted line corresponds to the HCM (with
hyperons).} \label{fig_6}
\end{figure}

Fig. \ref{fig_6} represents the baryon density profiles $\rho_B(r)$
predicted by the NJL model as well as by the HCM (with hyperons). In
all cases the results are shown for the respective maximum--mass stars.
The profile of a self--bound quark star at $\xi = 0$ has a step--like
behavior at the star's surface at radius $r \simeq 8$ km. Here
the density jumps from $\rho_B \simeq 2.6 \, \rho_0$ to a much lower value
corresponding to the crust density, or to zero if the crust is
neglected.  For $\xi = 0.5$ the corresponding EoS has a phase
transition at a finite (nonzero) pressure (see Sect. \ref{sec2c}). One
can see that in this case the density in the mixed--phase region
changes quite steeply but continuously.  For $\xi =1$ no phase
transition occurs and the baryon density decreases much smother
resulting in a bigger radius. The density profile of a maximum--mass
star predicted by the HCM starts at a much lower central baryon
density,
$\rho_c\simeq 6.5\,\rho_0$\,, but extends to a larger
radius. Comparing these results we conclude again that with increasing
$\xi$ the quark stars become more and more similar to neutron stars.

\subsection{Strangeness content}
\begin{figure}[ht]
\centerline{\mbox{
\epsfxsize=12cm\epsffile{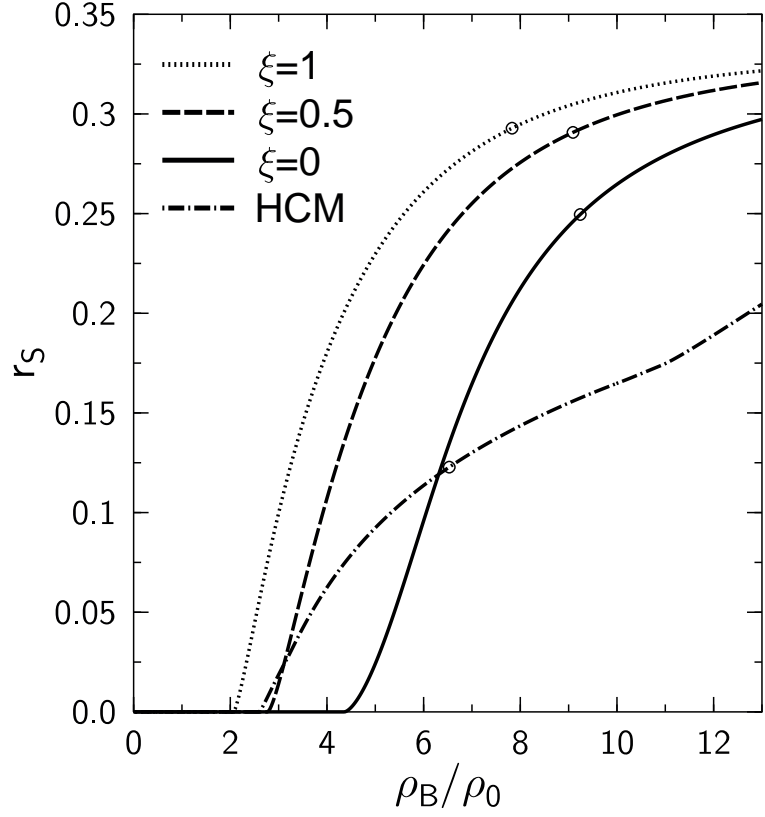}}}
\caption{Strangeness fraction $r_s$ versus the baryon density
$\rho_B$ predicted by the NJL model for three different values of the
parameter $\xi$\,. The dashed--dotted line shows the results of the HCM
(with hyperons). The circles correspond to the values of $r_s$ at the
center of the stars with maximum masses.} \label{fig_7}
\end{figure}

In this section we discuss the strangeness content of the quark matter
as predicted by the NJL model. In Fig. \ref{fig_7} the strangeness
fraction $r_s$ (see Eq. (\ref{rs})) is shown as a function of the
baryon density.  One can see that the threshold density for the
appearance of strange quarks depends significantly on the parameter
$\xi$ and diminishes with it.  At $\rho_B \goo 6 \, \rho_0$ the
effect of the vector coupling on $r_s$ becomes less important.
For all $\xi$ the strangeness fraction tends to $1/3$ that simply means
that dense quark matter is composed of equal numbers of up, down and
strange quarks. This is clear since the difference in bare masses of
light and strange quarks becomes unimportant at high densities and the
SU(3)--flavour symmetry is effectively restored. Similar to
neutron stars, stable quark stars should have smaller central densities
than the one corresponding to the maximum--mass star. This stability
condition gives a maximum value for the strangeness fraction
$r_s^{{\rm max}}$
in the star's center, indicated by open dots in Fig.
\ref{fig_7}. These values differ slightly for different $\xi$ values.
Indeed, $r_s^{{\rm max}}$ increases from $0.25$ to $0.29$ when $\xi$
changes from $0$ to $1$. In neutron star models dealing with
charge--neutral hadronic matter the strangeness fraction due to
formation of hyperons \cite{SM96,Hana00} is typically smaller than in
the quark stars. As one can see in Fig.~\ref{fig_7},
a much smaller value $r_s^{{\rm max}} \simeq 0.12$\,, is predicted by
the HCM.

\begin{figure}[ht]
\centerline{\mbox{
\epsfxsize=12cm\epsffile{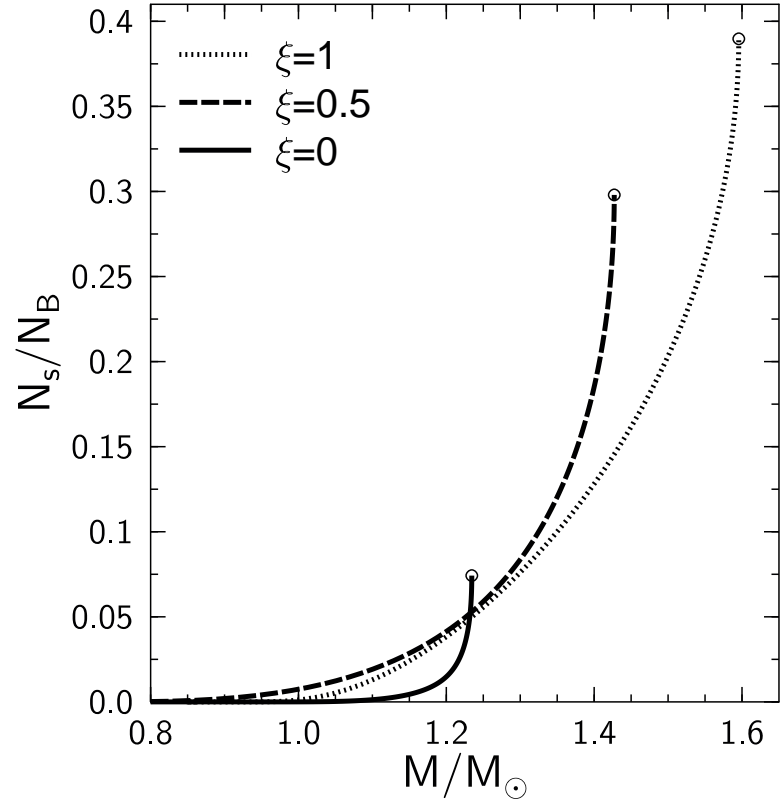}}}
\caption{The ratio of total number of strange quarks $N_s$ to the total
baryon number $N_B$ as function of the gravitational mass $M$ of the
quark stars predicted by the NJL model
for three values of the parameter $\xi$\,.} \label{fig_8}
\end{figure}

It is instructive to calculate the total strangeness content of a star.
It is obtained by integrating the strange number density $\rho_s$ over
the star's volume
\bel{Ns}
N_s = \int \sqrt{-g} \, \rho_s \, u^0 \, dV  = 4 \, \pi \int_0^R r^2
\frac{\rho_s}{\sqrt{1 - 2\hspace*{0.1em}m(r)/r}} \, dr \quad ,
\ee
where $u^0$ is the time component of the 4--velocity of matter and
$m(r)$ is the gravitational mass at the radius $r$\,.  The ratio of the
total strangeness number $N_s$ to the total baryon number $N_B$ is
shown in Fig.~\ref{fig_8} for different $\xi$ values. Due to the low
central density the low--mass stars ($M \loo M_\odot$) have a
negligible strangeness content for all $\xi$. For larger masses
$N_s/N_B$ is strongly sensitive to $\xi$. For the maximum--mass
stars the value of $N_s/N_B$ changes from $0.07$ to $0.39$ when $\xi$
increases from $0$ to $1$\,.  This difference originates from the
different density dependence of the strangeness fraction $r_s$ (see
Fig. \ref{fig_7}) in combination with the baryon density profiles of
the maximum--mass stars.  The stars with higher $\xi$ contain more
strange particles due to a lower threshold in density. The quark stars
with large strangeness content, say $N_s/N_B > 0.3$\, can be named
strange quark stars.

\section{Conclusions}
In this paper the NJL model is used to construct the EoS of cold
$\beta$--equilibrium quark matter and to study the structure of compact
stellar objects. It was found that the EoS depends strongly on the
relative strength of the vector and scalar interaction
$\xi = G_V/G_S$. When repulsive vector interaction is small
($\xi < 0.4$) the matter is self--bound, i.e.
there is a zero pressure point at a finite baryon density. At higher
values of $\xi$ (\mbox{$0.4 < \xi < 0.65$}) the EoS has no
zero--pressure points but still has a first order phase transition at
non--vanishing pressure. The properties of the mixed phase were found
by using the Gibbs conditions for the case of two independent chemical
potentials \cite{Gle92} and assuming global charge neutrality. We have
calculated the mass--density and mass--radius curves, the density
profiles and strangeness distribution in such stars. It is found that
maximum mass grows from $1.23 \, M_\odot$ to $1.6 \, M_\odot$ when
$\xi$ increases from $0$ to $1$. In the case of self--bound matter the
baryon density at the star's surface varies very rapidly from a high
value, corresponding to the zero--pressure point in the EoS, to nearly
zero density. The radii of such stars could be quite small even if the
crust is added (the minimum radius in this case is about $7$ km). At
$\xi \sim 1$ the quark star global properties (maximum mass
$\simeq 1.6 \, M_\odot$, radius $\simeq 11$ km) are similar to those
for neutron stars.  However, the quark stars have much higher
strangeness content as compared with stars composed of hadronic matter
with hyperons. These quark stars near their maximum masses can be aptly
dubbed strange quark stars.

In the future we are planing to construct a more realistic EoS when
hadronic degrees of freedom will be included at low densities.

\section*{Acknowledgments}
The authors are thankful to J. Schaffner--Bielich for helpful discussions.
This work has been supported by the RFBR Grant No. 00--15--96590,
the Alexander von Humboldt Foundation, the Graduiertenkolleg ``Experimentelle
und Theoretische Schwerionenphysik'', GSI, BMBF, DFG, and the
Hessische Landesgraduiertenf\"orderung.



\end{document}